\documentclass[default,iicol]{sn-jnl}
\usepackage{bm}
\usepackage{xcolor}
\newcommand{\beginsupplement}{%
        \setcounter{table}{0}
        \renewcommand{\thetable}{S\arabic{table}}%
        \setcounter{figure}{0}
        \renewcommand{\thefigure}{S\arabic{figure}}%
}
\usepackage{braket}
\usepackage{tabularx}
\usepackage{siunitx}
\DeclareSIUnit\bar{bar}
\usepackage{upgreek}
\usepackage{xfrac}

\newcommand{\Beplus}{\ensuremath{{^9}{\rm Be}^{+} \,}}
\newcommand{\up}{\ket{\uparrow}}
\newcommand{\down}{\ket{\downarrow}}
\newcommand{\mJ}{m_{\text{J}}}
\newcommand{\mI}{m_{\text{I}}}
\newcommand{\smanifold}{$2\text{s}~{}^{2}\text{S}_{\sfrac{1}{2}}$}
\newcommand{\phighmanifold}{$2\text{p}~{}^{2}\text{P}_{\sfrac{3}{2}}$}

\usepackage{graphicx}%
\usepackage{multirow}%
\usepackage{amsmath,amssymb,amsfonts}%
\usepackage{amsthm}%
\usepackage{mathrsfs}%
\usepackage[title]{appendix}%
\usepackage{xcolor}%
\usepackage{textcomp}%
\usepackage{manyfoot}%
\usepackage{booktabs}%
\usepackage{algorithm}%
\usepackage{algorithmicx}%
\usepackage{algpseudocode}%
\usepackage{listings}
\usepackage{braket}
\usepackage{hyperref}

\usepackage[]{multibib}
\newcites{Supp}{Methods References}
\begin{document}

\title[Penning micro-trap for quantum computing]{Penning micro-trap for quantum computing}

\author*[1,2]{\fnm{Shreyans} \sur{Jain}}\email{sjain@phys.ethz.ch}
\equalcont{These authors contributed equally to this work.}
\author[1,2]{\fnm{Tobias} \sur{S\"agesser}}
\equalcont{These authors contributed equally to this work.}
\author[1,2]{\fnm{Pavel} \sur{Hrmo}}
\author[1]{\fnm{Celeste} \sur{Torkzaban}}
\author[1,2]{\fnm{Martin} \sur{Stadler}}
\author[1,2]{\fnm{Robin} \sur{Oswald}}
\author[1]{\fnm{Chris} \sur{Axline}}
\author[3,4]{\fnm{Amado} \sur{Bautista-Salvador}}
\author[3,4]{\fnm{Christian} \sur{Ospelkaus}}
\author[1,2]{\fnm{Daniel} \sur{Kienzler}}
\author[1,2]{\fnm{Jonathan} \sur{Home}}
\affil[1]{\orgdiv{Department of Physics}, \orgname{ETH Z{\"u}rich}, \orgaddress{\city{Z{\"u}rich}, \country{Switzerland}}}
\affil[2]{\orgdiv{Quantum Center}, \orgname{ETH Z{\"u}rich}, \orgaddress{\city{Z{\"u}rich}, \country{Switzerland}}}
\affil[3]{\orgdiv{Institut f{\"u}r Quantenoptik}, \orgname{Leibniz Universit{\"a}t Hannover}, \orgaddress{\city{Hannover}, \country{Germany}}}
\affil[4]{\orgname{Physikalisch-Technische Bundesanstalt}, \orgaddress{\city{Braunschweig}, \country{Germany}}}


\abstract{
Trapped ions in radio-frequency traps are among the leading approaches for realizing quantum computers, due to high-fidelity quantum gates and long coherence times \cite{Ballance2016,Clark2021,Srinivas2021}. However, the use of radio-frequencies presents a number of challenges to 
scaling, including requiring compatibility of chips with high voltages \cite{Brown2021}, managing power dissipation \cite{Malinowski2023} and restricting transport and placement of ions \cite{palani_high-fidelity_2023}. 
Here we realize a micro-fabricated Penning ion trap which removes these restrictions by replacing the radio-frequency field with a 3 T magnetic field.
We demonstrate full quantum control of an ion in this setting, as well as the ability to transport the ion arbitrarily in the trapping plane above the chip. This unique feature of the Penning micro-trap approach opens up a modification of the Quantum CCD architecture with improved connectivity and flexibility, facilitating the realization of large-scale trapped-ion quantum computing, quantum simulation and quantum sensing.}

\maketitle

Trapped atomic ions are among the most advanced technologies for realizing quantum computation and quantum simulation, based on a combination of  high-fidelity quantum gates \cite{Ballance2016,Clark2021,Srinivas2021} and long coherence times  \cite{Wang2021}. These have been used to realize small-scale quantum algorithms  and quantum error correction protocols. However, scaling the system size to support orders-of-magnitude more qubits \cite{Gidney2021,Alexeev2021} appears highly challenging \cite{Pogorelov2021,Cetina2022,Kranzl2022,Pino2021}. One of the primary paths to scaling is the Quantum CCD (QCCD) architecture, which involves arrays of trapping zones between which ions are shuttled during algorithms \cite{Kielpinski2002a, Home2009, Pino2021, Ryan-Anderson2021}. However, challenges arise due to the intrinsic nature of the radio-frequency (rf) fields, which require specialized junctions for 2-dimensional (2-d) connectivity of different regions of the trap. Although successful demonstrations of junctions have been performed, these require dedicated large-footprint regions of the chip which limits trap density \cite{zhangOptimizationImplementationSurfaceelectrode2022,Blakestad2009,Moehring2011,Shu2014,Burton2023}. This adds to several other undesirable features of the rf drive which make micro-trap arrays difficult to operate \cite{palani_high-fidelity_2023}, including significant power dissipation due to the currents flowing in the electrodes, and the need to co-align the rf and static potentials of the trap to minimize micromotion, which affects gate operations \cite{Berkeland1998, Jain2020}. Power dissipation is likely to be a very severe constraint in trap arrays of more than 100 sites \cite{Malinowski2023, Jain2020}.

\par 
An alternative to rf electric fields for radial confinement is to use a Penning trap where only static electric and magnetic fields are employed, which is an extremely attractive feature for scaling due to the lack of power dissipation and geometrical restrictions on placement of ions \cite{Hellwig2010, Jain2020}. Penning traps are a well established tool for precision spectroscopy with small numbers of ions \cite{Ahmadi2017,Ulmer2015,Hanneke2008,DiSciacca2012}, while quantum simulations and quantum control have been demonstrated in crystals of more than 100 ions \cite{Britton2012,Bohnet2016,Stutter2018}. However, the single trap site used in these approaches does not provide the flexibility and scalability necessary for large-scale quantum computing. 

Invoking the approach of the QCCD architecture, a scalable approach can be envisioned to be the Penning QCCD, in which a micro-fabricated electrode structure allows trapping of ions at many individual trapping sites, which can be actively reconfigured during the algorithm by changing the electric potential. Beyond the static arrays considered in previous work \cite{Jain2020, stahl_planar_2005}, we here envisage that ions in separated sites are brought close to each other to utilize the Coulomb interaction for two-qubit gate protocols implemented through applied laser or microwave fields \cite{leibfried_experimental_2003, Wilson2014}, before being transported to additional locations for further operations. A major advantage of this approach is that transport of ions can be performed in three dimensions almost arbitrarily without the need for specialized junctions, allowing flexible and deterministic reconfiguration of the array with low spatial overhead. 

In this article, we demonstrate the fundamental building block of such an array by trapping a single ion in a cryogenic micro-fabricated surface-electrode Penning trap. We demonstrate quantum control of its spin and motional degrees of freedom, and measure a heating rate lower than in any comparably sized rf trap. We use this system to  demonstrate flexible  2-d transport of ions above the electrode plane with  negligible heating of the motional state. This provides a key ingredient for scaling based on the Penning ion-trap QCCD architecture.

\begin{figure*}[ht]
\centering
\resizebox{\textwidth}{!}{\includegraphics{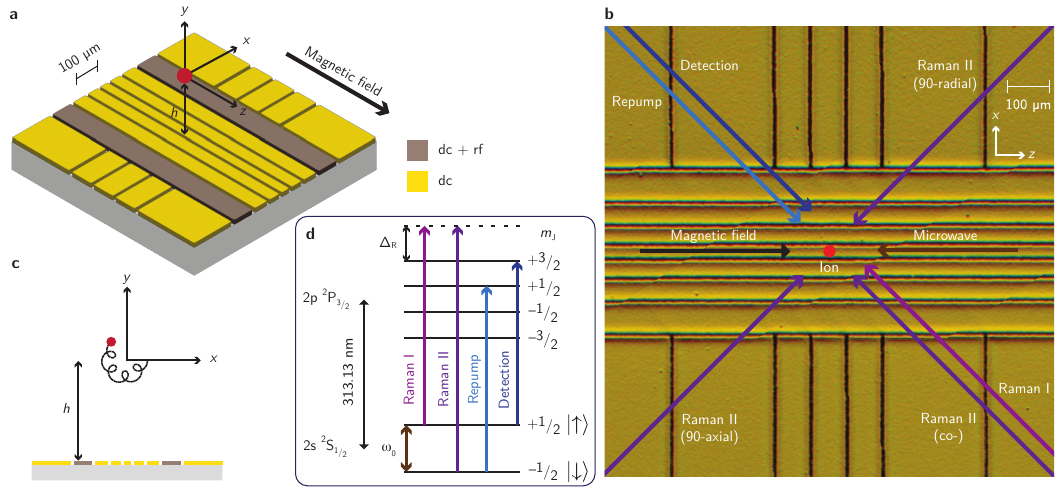}}
\caption{\label{fig:trapbeams}\textbf{Surface-electrode Penning trap.} 
\textbf{a.} Schematic showing the middle section of the micro-fabricated surface-electrode trap. The trap chip is embedded in a uniform magnetic field along the $z$-axis, and the application of dc voltages on the electrodes leads to 3-d confinement of the ion at a height $h \simeq 152~\upmu\mathrm{m}$ above the surface. Electrodes labelled ``dc + rf'' are used for coupling the radial modes during Doppler cooling. 
\textbf{b.} Micrographic image of the trap chip, with an overlay of the direction of the laser beams (all near 313 nm) and microwave radiation (near $\omega_0 \simeq 2 \pi \times 83.2$ GHz) required for manipulating the spin and motion of the ion. All laser beams run parallel to the surface of the trap and are switched on or off using acousto-optic modulators (AOMs), while the microwave radiation is delivered to the ion via a horn antenna close to the chip. 
\textbf{c.} Epicyclic motion of the ion in the radial plane ($x$-$y$) resulting from the sum of the two circular eigenmodes, the cylotron and the magnetron modes.
\textbf{d.} Electronic structure of the $\Beplus$ ion, with the relevant transitions used for coherent and incoherent operations on the ion. Only the  levels with nuclear spin $\mI = +\sfrac{3}{2}$ are shown. The virtual level (dashed line) used for Raman excitation is detuned $\Delta_{\mathrm{R}} \simeq + 2 \pi \times 150$ GHz from the \phighmanifold~$\ket{\mI = +\sfrac{3}{2}, \mJ = +\sfrac{3}{2}}$~state.}
\end{figure*}

The experimental setup involves a single beryllium ($\Beplus$) ion confined using a static quadrupolar electric potential generated by applying voltages to the electrodes of a surface-electrode trap with geometry shown in (fig. \ref{fig:trapbeams} a-c). We use a radially symmetric potential $V(x,y,z) = m \omega_z^2 (z^2 - (x^2 + y^2)/2)/(2e)$, centred at a position  $152~\upmu$m above the chip surface. 
The trap is embedded in a homogeneous magnetic field aligned along the $z$-axis with a magnitude of $B \simeq 3$ T, supplied by a superconducting magnet. The trap assembly is placed in a cryogenic, ultra-high vacuum chamber which fits inside the magnet bore, with the aim of reducing background-gas collisions  and motional heating. Using a laser at 235 nm, we load the trap by resonance-enhanced multiphoton ionization of neutral atoms produced from either a resistively heated oven or an ablation source \cite{Lo2014}. We regularly trap single ions for more than a day, with the primary loss mechanism being related to user interference. Further details about the apparatus can be found in the Methods.

The 3-d motion of an ion in a Penning trap can be described as a sum of three harmonic eigenmodes. The  axial motion along $z$ is a simple harmonic oscillator at frequency $\omega_z$. The radial motion is composed of modified-cyclotron ($\omega_+$) and magnetron ($\omega_-$) components, with frequencies $\omega_{\pm} = \omega_c/2  \pm \Omega$ where $\Omega = \sqrt{\omega_c^2 - 2 \omega_z^2}/2 $\cite{Brown1986} and $\omega_c = e B / m \simeq 2\pi \times$~5.12 MHz is the bare cyclotron frequency. Voltage control over the dc electrodes of the trap allows the axial frequency to be set to any value up to the stability limit, $\omega_z \leq \omega_c/\sqrt{2} \simeq 2\pi \times$~3.62 MHz. This corresponds to a range $ 0 \leq \omega_- \leq 2 \pi \times 2.56$~MHz and $2 \pi \times2.56$~MHz~$\leq \omega_+ \leq 2 \pi \times5.12$ MHz for the magnetron and modified-cyclotron modes respectively. Doppler cooling of the magnetron mode, which has a negative total energy, is achieved using a weak axialization rf quadrupolar electric field ($\leq$~60 mV peak-to-peak voltage on the electrodes) at the bare cyclotron frequency, which resonantly couples the magnetron and modified-cyclotron motions  \cite{Powell2002,Hrmo2019}. For the wiring configuration used in this work, the null of the rf field is produced at a height $h \simeq 152 \upmu$m above the electrode-plane. Aligning the null of the dc (trapping) field to the rf null is beneficial because it reduces driven radial motion at the axialization frequency; nevertheless, we find that Doppler cooling works with a relative displacement of tens of micrometres between the dc and rf nulls, albeit with lower efficiency. The rf field is required only during Doppler cooling, and not, for instance, during coherent operations on the spin or motion of the ion. All measurements in this work are taken at an axial frequency $\omega_z \simeq 2 \pi \times 2.5$ MHz, unless stated otherwise. The corresponding radial frequencies are $\omega_+ \simeq 2 \pi \times 4.41$ MHz and $\omega_- \simeq 2 \pi \times 0.71$ MHz. 

Fig. \ref{fig:trapbeams} d shows the electronic structure of the beryllium ion along with the transitions relevant to this work. We use an electron spin qubit (consisting of the $\up \equiv \ket{\mI = +\sfrac{3}{2}, \mJ = +\sfrac{1}{2}}$ and $\down \equiv \ket{\mI = +\sfrac{3}{2}, \mJ = -\sfrac{1}{2}}$ eigenstates within the \smanifold~ground-state manifold) which in the high field is almost decoupled from the nuclear spin. The qubit frequency is $\omega_0 \simeq 2 \pi \times 83.2$ GHz. Doppler cooling is performed using the detection laser red-detuned from the (bright) $\up \leftrightarrow$ \phighmanifold~$ \ket{\mI = +\sfrac{3}{2}, \mJ = +\sfrac{3}{2}}$ cycling transition, while an additional repump laser optically pumps population from the (dark) $\down$ level to the higher energy $\up$ level via the fast-decaying \phighmanifold~$\ket{\mI = +\sfrac{3}{2}, \mJ = +\sfrac{1}{2}}$ excited state. State-dependent fluorescence with the detection laser allows for discrimination between the two qubit states based on photon counts collected on a photomultiplier tube using an imaging system which uses a 0.55 NA Schwarzschild objective. The fluorescence can also be sent to an electron-multiplying CCD (EMCCD) camera.

\par Coherent operations on the spin and motional degrees of freedom of the ion are performed either using stimulated Raman transitions with a pair of lasers tuned to $150$ GHz above the \phighmanifold~$\ket{\mI = +\sfrac{3}{2}, \mJ = +\sfrac{3}{2}}$~state, or using a microwave field. The former requires the use of two 313 nm lasers phase-locked at the qubit frequency which we achieve using the method outlined in reference \cite{Mielke2021}. By choosing different orientations of Raman laser paths, we can address the radial or axial motions, or implement  single qubit rotations using a  co-propagating Raman beam pair.

The qubit transition has a sensitivity of 28 GHz/T to the magnetic field, meaning the phase-coherence of our qubit is susceptible to temporal fluctuations or spatial gradients of the field across the extent of the ion's motion. Using Ramsey spectroscopy, we measure a coherence time of 1.9(2) ms with the Raman beams. Similar values are measured with the microwave field, indicating that laser phase-noise from beam path fluctuations or imperfect phase-locking does not significantly contribute to dephasing. The nature of the noise appears to be slow on the timescale($\sim 1~\mathrm{ms}-10$~ms) of a single experimental shot consisting of cooling, probing and detection, and the fringe contrast decay follows a Gaussian curve. We notice that the coherence is reduced if vibrations induced by the cryocoolers used to cool the magnet and the vacuum apparatus are not well decoupled from the experimental setup. Further characterization of the magnetic field noise is performed via applying different orders of the Uhrig dynamical decoupling sequence \cite{Uhrig2007,Biercuk2009}, with resulting  extracted coherence time from the measurements being 3.2(1) ms, 
5.8(3) ms and 8.0(7) ms for orders 1, 3 and 5 respectively. Data on spin-dephasing are presented in Extended Data Figure \ref{fig:SpinCoh}. 

A combination of the Doppler cooling and repump lasers prepares the ion in the $\up$ electronic state and a thermal distribution of motional Fock states. After Doppler cooling using the axialization technique, we measure mean occupations of 
$\{ \bar{n}_+ , \bar{n}_-, \bar{n}_z \} = \{ 6.7(4), 9.9(6), 4.4(1) \}$
 via sideband spectroscopy on the first four red and blue sidebands \cite{Hrmo2019}.
Pulses of continuous sideband cooling \cite{Stutter2018,Hrmo2019} are subsequently performed by alternatively driving the first and third blue (red) sidebands of a positive (negative) energy motional mode while simultaneously repumping the spin state to the bright state. The 3-d ground state can be prepared by applying this sequence for each of the three modes in succession. The use of the third sideband is motivated by the high Lamb-Dicke parameters of approximately 0.4 in our system \cite{Chen2017, Joshi2019}.  After a total time of 60~ms of cooling we probe the temperature using sideband spectroscopy on the first blue and red sidebands \cite{Monroe1995}. Assuming thermal distributions, we  measure 
$\{\bar{n}_+, \bar{n}_-, \bar{n}_z\} = \{0.05(1), 0.03(2), 0.007(3)\}$. We have achieved similar performance of the ground-state cooling at all trap frequencies probed to date. The long duration of the sideband cooling sequence stems from the large (estimated as $80~\upmu$m) Gaussian beam radius of the Raman beams each with a power in the range 2 mW - 6 mW, leading to a Rabi frequency $\Omega_0\simeq2\pi\times8$ kHz, which corresponds to pi-times of approximately $62~\upmu$s, $145~\upmu$s and $2000~\upmu$s for the ground state carrier, first and third sidebands respectively at $\omega_z=2\pi\times 2.5$ MHz.\\

\begin{figure*}[ht]
\centering
\resizebox{\textwidth}{!}{\includegraphics{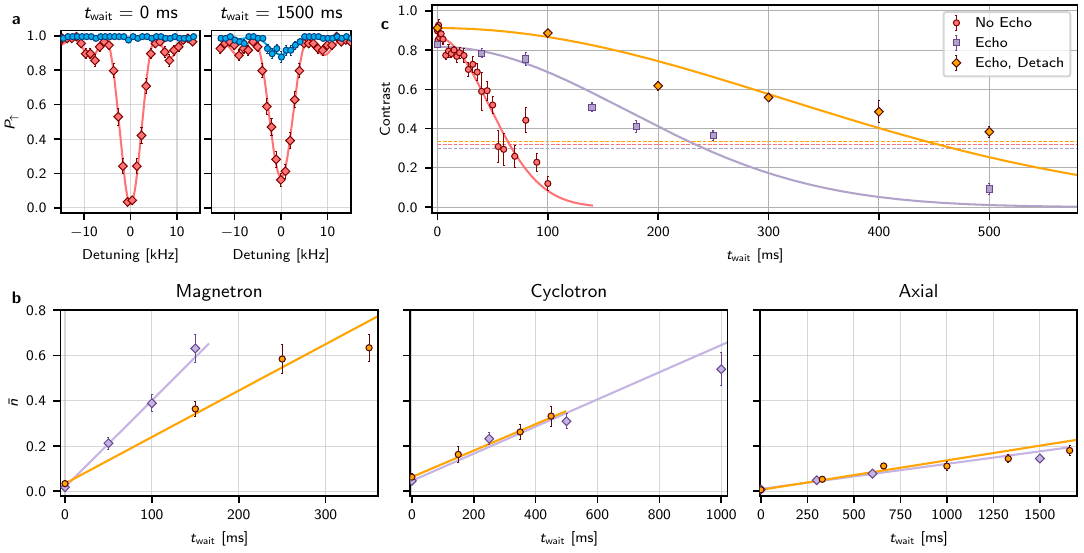}}
\caption{\label{fig:SBC_heating}\textbf{Motional coherence.} \textbf{a.} Bright state population, $P_{\uparrow}$, measured after applying the first red or blue axial sideband probe-pulse to the sideband-cooled ion. Since the bright state, $\ket{\uparrow}$, has a higher energy than the dark state, $\ket{\downarrow}$, the blue sideband can not be driven when the ion is in the ground state of the axial mode. \textbf{b.} Average phonon number $\bar{n}$ calculated using the sideband-ratio method \cite{Monroe1995} for all three modes as a function of increasing $t_{\mathrm{wait}}$. Points in purple and orange indicate data taken with the trap connected and detached, respectively. The heating rates are extracted from the slopes of the linear fits. \textbf{c.} Motional dephasing of the axial mode observed via Ramsey spectroscopy. Points in purple indicate data taken with an echo pulse in the sequence. Points in orange indicate data taken with an echo pulse, where additionally the trap was detached between Doppler cooling and the detection pulse.  While the dataset with the voltage sources detached is taken at $\omega_z \simeq 2 \pi \times 2.5$ MHz, the two data series with the trap attached are taken at an axial mode frequency, $\omega_z \simeq 2 \pi \times 3.1$ MHz. The dashed lines show the $1/e$ line normalized to the Gaussian fits. All error bars in this figure indicate the standard error.}
\end{figure*}

Trapped-ion quantum computing utilizes the collective motion of the ions for multi-qubit gates, and thus requires the motional degree of freedom to retain coherence over the timescale of the operation \cite{Sorensen2000, leibfried_experimental_2003}. One contribution to decoherence comes from motional heating due to fluctuations in the electric field at frequencies close to the oscillation frequencies of the ion. We measure this by inserting a variable-length delay $t_{\mathrm{wait}}$ between the end of sideband cooling and the temperature probe. As shown in figure \ref{fig:SBC_heating}, we observe motional heating rates 
$\{ \dot{\bar{n}}_+, \dot{\bar{n}}_-, \dot{\bar{n}}_z  \}= \{0.49(5)~\mathrm{s}^{-1}, 3.8(1)~\mathrm{s}^{-1}, 0.088(9)~\mathrm{s}^{-1} \}$. The corresponding electric-field spectral noise density for the axial mode, $S_E  = 4\hbar m \omega_z \dot{\bar{n}}_z / e^2 = 3.4(3)\times {10}^{-16}~\mathrm{V}^2 \mathrm{m}^{-2} \mathrm{Hz}^{-1}$, is lower than any comparable measurement in a trap of similar size \cite{Lakhmanskiy2019,brownnuttIontrapMeasurementsElectricfield2015}. As detailed in the Methods, we are able to trap ions in our setup with the trap electrodes detached from any external supply voltage except during Doppler cooling, which requires the axialization signal to pass to the trap. Using this method, we measure heating rates $\dot{\bar{n}}_z =  0.10(1)~\mathrm{s}^{-1}$ and $\dot{\bar{n}}_+ = 0.58(2) ~\mathrm{s}^{-1}$ for the axial and cyclotron modes respectively, while the rate for the lower-frequency magnetron mode drops to $\dot{\bar{n}}_- = 1.8(3) ~\mathrm{s}^{-1}$. This reduction suggests that external electrical noise contributes to the higher magnetron heating rate in the earlier measurements.

Motional state dephasing was measured using Ramsey spectroscopy, involving setting up a superposition $\ket{\uparrow} \left(\ket{0}_z + \ket{1}_z \right) / \sqrt{2}$ of the first two Fock states of the axial mode (here $\omega_z \simeq 2 \pi \times 3.1$ MHz) using a combination of carrier and sideband pulses \cite{Turchette2000}. Following a variable wait time we reverse the preparation sequence with a shifted phase. The resulting decay of the Ramsey contrast shown in figure \ref{fig:SBC_heating}c is much faster than what would be expected from the heating rate. The decay is roughly Gaussian in form with a $1/\mathrm{e}$-coherence time of 66(5) ms. Inserting an echo pulse in the Ramsey sequence extends the coherence time to 240(20) ms, which indicates low-frequency noise components dominating the bare Ramsey coherence. Further improvement of the echo coherence time to 440(50) ms is observed when the trap electrodes are detached from external voltage sources between the conclusion of Doppler cooling and the start of the detection pulse, where again the axialization signal is beneficial. The data with the voltage sources detached is taken at $\omega_z \simeq 2 \pi \times 2.5$ MHz.

\par
\begin{figure}[!ht]
\resizebox{\columnwidth}{!}{\includegraphics{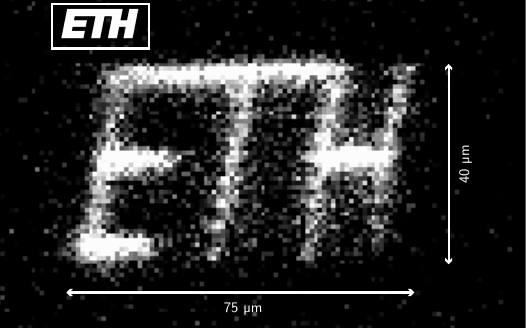}}
\caption{\label{fig:2dtransport}\textbf{2-d transport demonstration.} A single ion is transported adiabatically in the $x$-$z$ plane (normal to the imaging optical axis). The ion is illuminated for 500 \si{\micro\second} at a total of 58 positions, here defined by the ETH Z\"urich logo (see inset for reference image). The red circle indicates the initial position where the ion is Doppler cooled. The ion is moved across a region spanning approximately 40 \si{\micro\metre} and 75  \si{\micro\metre} along the $x$ (radial) and $z$ (axial) directions respectively. The sequence is repeated 172 times to accumulate the image.}
\end{figure}
\begin{figure*}[ht]
\resizebox{\textwidth}{!}{\includegraphics{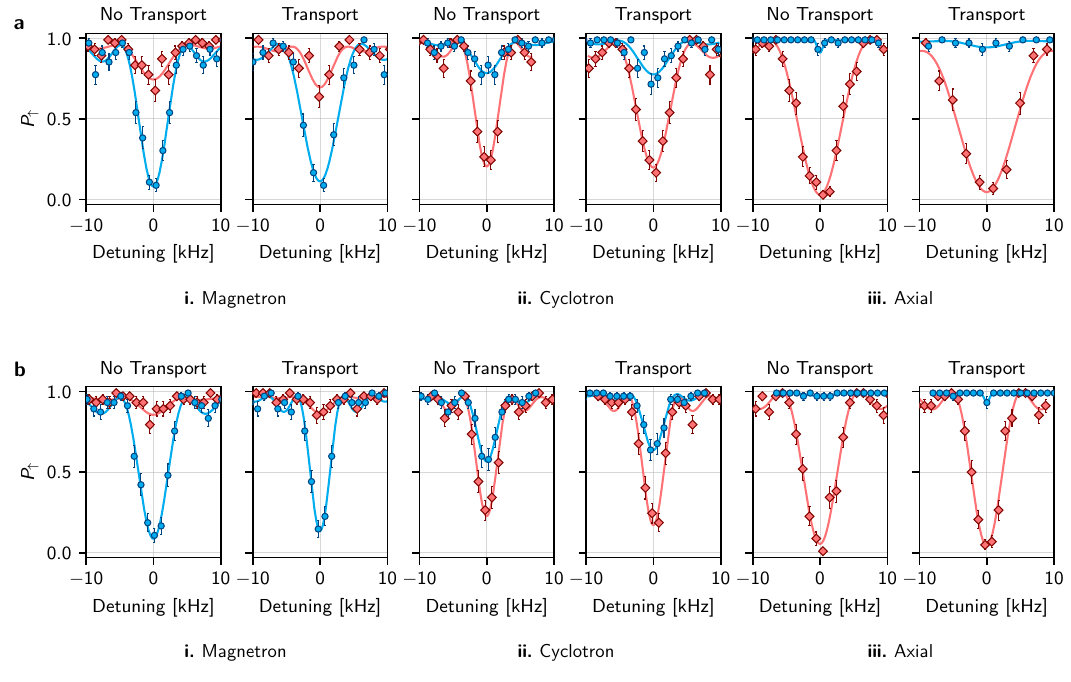}}
\caption{\label{fig:TransHeat} \textbf{Heating during transport}. Bright-state population $P_{\uparrow}$ measured after applying the first red or blue axial sideband probe-pulse before and after a number of shuttling events. In panel \textbf{a}, the ion is moved along the $z$-axis (axial direction), while in panel \textbf{b}, the ion is moved along the $x$-axis (in-plane radial direction) - in both cases, the distance moved is approximately $30~\upmu$m. For the axial mode, the transport sequence between the initial and final locations is repeated 10 times before the temperature is probed. A single one-way transport sequence has a duration of approximately $300~\upmu$s. The temperature is then compared to the case where the ion is held at the initial position for an equivalent total waiting time; this case is labelled as `No Transport' in the figure. In the case of the magnetron and cyclotron modes, the transport-sequence is repeated 4 times instead. We see no appreciable change in the ion temperature when the transport time is increased. All error bars in this figure indicate the standard error.}
\end{figure*}

A critical component of the QCCD architecture \cite{Kielpinski2002a} is ion transport. 
We demonstrate that the Penning trap approach allows us to perform this flexibly in two dimensions by adiabatically transporting a single ion, and observing it at the new location. The ion is first Doppler cooled at the original location, and then transported in 4 ms to a second desired location along a direct trajectory. We then perform a \SI{500}{\micro\second} detection pulse without applying axialization, and collect the ion fluorescence on an EMCCD camera. The exposure of the camera is limited to the time-window defined by the detection pulse. The lack of axialization is important when the ion is sufficiently far from the rf null in order to minimize radial excitation due to micromotion, and subsequently produce enough fluorescence during the detection window. The ion is then returned to the initial location. Figure \ref{fig:2dtransport} shows a result where we have drawn the first letters of the ETH~Z{\"u}rich logo. The image quality and maximum canvas size are only limited by the point-spread function and field of view of our imaging system, as well as the spatial extent of the detection laser beam, and not by any property of the transport. Reliable transport to a set location and back has been performed up to 250 microns. By probing ion temperatures after transport using sideband themometry (see Extended Data Figure \ref{fig:TransHeat}), we have observed no evidence of motional excitation from transport compared to the natural heating expected over the duration of the transport. This contrasts with earlier non-adiabatic radial transport of ensembles of ions in Penning traps, where a good fraction of the ions were lost in each transport \cite{Crick2010a}.  

This work marks a starting point for quantum computing and simulation in micro-scale Penning trap 2-d arrays. The major next steps are to operate with multiple sites of such an array, which will require optimization of the loading while keeping ions trapped in shallow potentials. This can be accomplished in the current trap with the appropriate wiring, but significant advantages could be gained by using a trap with a loading region and shuttling ions in to the micro-trap region. Multi-qubit gates could then be implemented following standard methods demonstrated in rf traps \cite{Wilson2014, Jain2020}.  Increased spin coherence times could be achieved through improvements to mechanical stability of the magnet, or in the longer term through the use of decoherence-free-subspaces, which were envisioned in the original QCCD proposals \cite{winelandExperimentalIssuesCoherent1998, Kielpinski2002a,ciracScalableQuantumComputer2000}. For scaling to large numbers of sites, it is likely that scalable approaches to light delivery will be required, which might necessitate switching to an ion species which is more amenable to integrated optics \cite{mehtaIntegratedOpticalMultiion2020, mehtaIntegratedOpticalAddressing2016, niffeneggerIntegratedMultiwavelengthControl2020, ivoryIntegratedOpticalAddressing2021a}. The use of advanced standard fabrication methods such as CMOS \cite{mehtaIonTrapsFabricated2014, auchterIndustriallyMicrofabricatedIon2022} is facilitated compared to rf traps by the lack of high-voltage rf signals. Compatibility with these technologies demands an evaluation of how close to the surface ions could be operated for quantum computing and will require in-depth studies of heating -  here an obvious next step is to sample electric field noise as a function of ion-electrode distance \cite{brownnuttIontrapMeasurementsElectricfield2015}. Unlike in rf traps, 3-d scans of electric field noise are possible in any Penning trap due to the flexibility of confinement from the uniform magnetic field. This flexibility of ion placement has advantages in many areas of ion trap physics, for instance in placing ions in anti-nodes of optical cavities \cite{Schupp2021}, or sampling field noise from surfaces of interest \cite{Hite2021,McKay2021}. We therefore expect that our work will open new avenues in sensing, computation, simulation and networking, allowing ion trap physics to break out beyond its current constraints. 

\bibliographystyle{unsrt}
\bibliography{main.bbl}

\newpage

\beginsupplement
\section*{Methods}

\subsection*{Spin-Coherence Data}
\begin{figure*}[ht]
\resizebox{\textwidth}{!}{\includegraphics{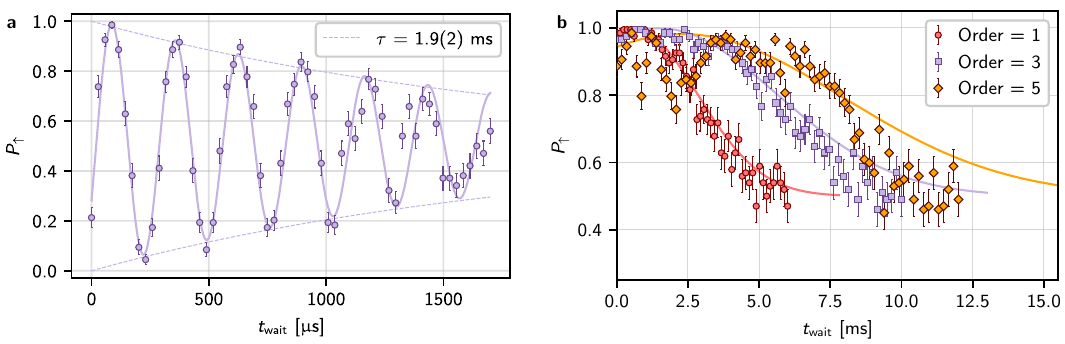}}
\caption{\label{fig:SpinCoh}\textbf{Spin coherence.} Population in the $\ket{\uparrow}$ state \textbf{a.} as a function of the wait time $t_{\mathrm{wait}}$ in the detuned Ramsey experiment, and \textbf{b.} for the first three odd orders of the Uhrig dynamical decoupling sequence performed using the Raman beams. In the Uhrig sequence, the phase of the final pulse is chosen to maximise population at 0 wait time. As there is always a $\pi$-pulse in the centre of the wait time, any static detuning offsets are cancelled and the population in the $\ket{\uparrow}$ state is proportional to the contrast. The coherence time is extracted from an empirical fit to an exponential function in \textbf{a.} and a Gaussian in \textbf{b.}.  All error bars in this figure indicate the standard error.}
\end{figure*}

The data for spin-dephasing measurements can be found in Extended Data Figure \ref{fig:SpinCoh}. The $N$-th order Uhrig sequence is performed by adding $N$ $\pi$-pulses centered at times
\begin{equation}
    \tau_n=t_{\text{wait}}\sin^2\left (\frac{n\pi}{2(N+1)}\right)
\end{equation}
between the two $\sfrac{\pi}{2}$-pulses.


\subsection*{Trap design and fabrication}
\par
The design of our surface-electrode Penning trap is based on the traditional 4-wire or 5-wire traps widely used in the rf trapped-ion community \citeSupp{Chiaverini2005}. Two configurations of arranging ions were envisaged during the design phase - a) a single trapping site that can be placed at a variable height above the surface, and b) two trap sites separated along the in-plane radial axis. 
In total the trap consists of 25 electrodes laid out on a planar surface and surrounded by a conducting ground plane. All electrodes are supplied with static voltages, while an individual rf signal can be applied to each of the middle 7 electrodes running parallel to the magnetic field, such that they also potentially serve as axialization electrodes - see fig. \ref{fig:trapbeams} b in the main text. The static voltages applied to the electrodes allow us to control the elements of the Hessian matrix of the trapping potential.
While typical rf traps might use two or three rf electrodes, the extra rf electrodes on our trap allow us to provide two nodal lines with variable separation in the rf axialization potential, or to produce a node at a variable height above the chip surface.
Experiments included in this article have been restricted to ions in a single trap site $152~\upmu$m above the trap surface, with only the outermost rf electrodes used for axialization. This height is determined using an analytical calculation and confirmed via an independent simulation based on the boundary element method.
\par
The trap was fabricated at PTB, Braunschweig, using a single layer processing method \citeSupp{Bautista-Salvador_2019} depositing gold electrodes via electroplating on a sapphire substrate. 
The process yields an electrode thickness in the range $10~\upmu$m to $12~\upmu$m and approximately $3~\upmu$m-wide gaps between the electrodes, which provides excellent electrical shielding of the substrate.

\subsection*{Cryogenic vacuum apparatus}

\begin{figure*}[ht]
\resizebox{\textwidth}{!}{\includegraphics{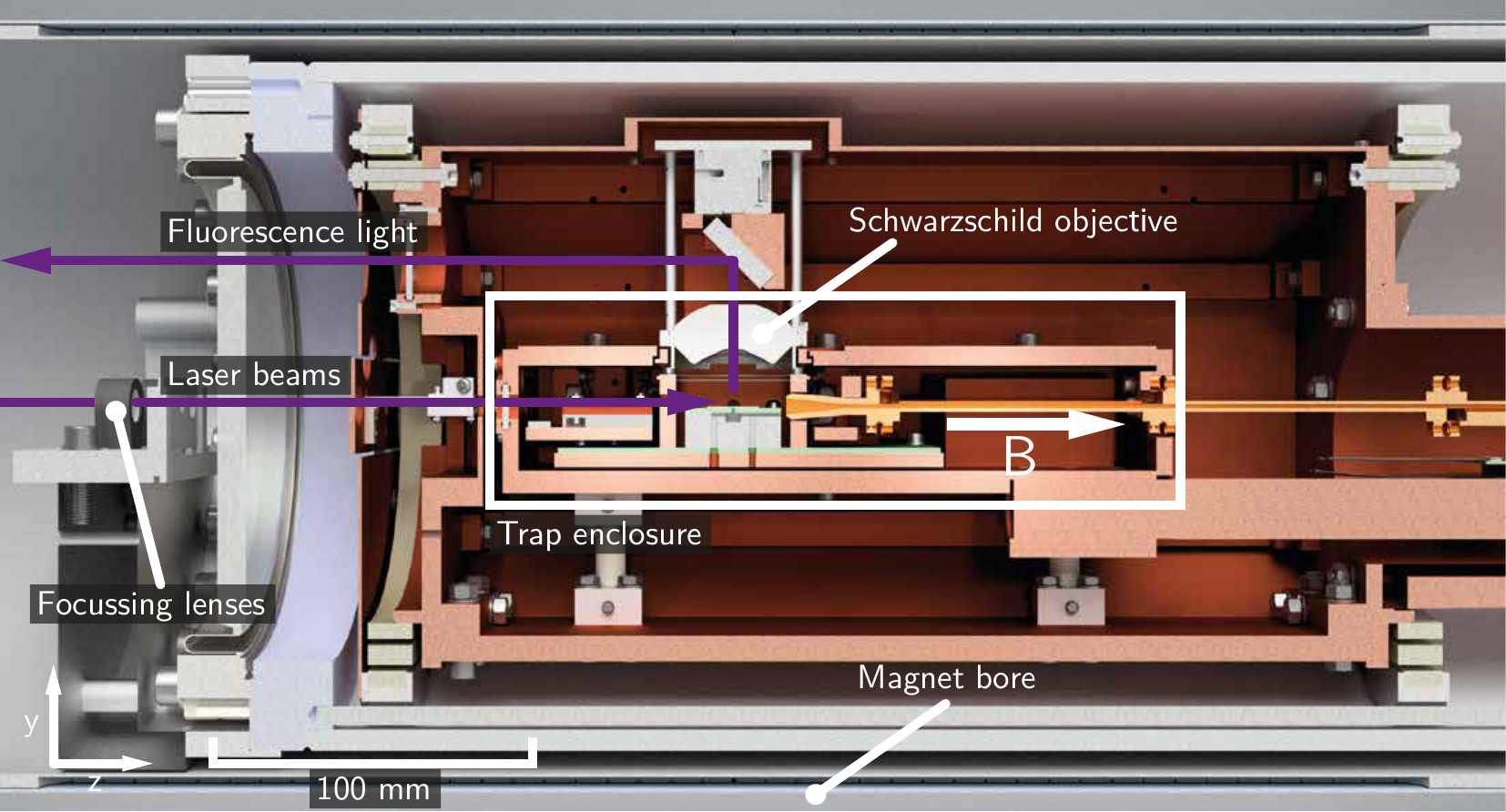}}
\caption{\label{fig:apparatus_overview}\textbf{Cryogenic vacuum apparatus.} Rendering of the cryogenic vacuum apparatus placed inside the magnet bore. Laser light is delivered along the bore to the trap enclosure, passing focusing lenses and a vacuum viewport. Ion fluorescence is collected using a Schwarzschild objective and directed out of the apparatus and the magnet bore using a mirror.}
\end{figure*}

\begin{figure}[ht]
\resizebox{\columnwidth}{!}{\includegraphics{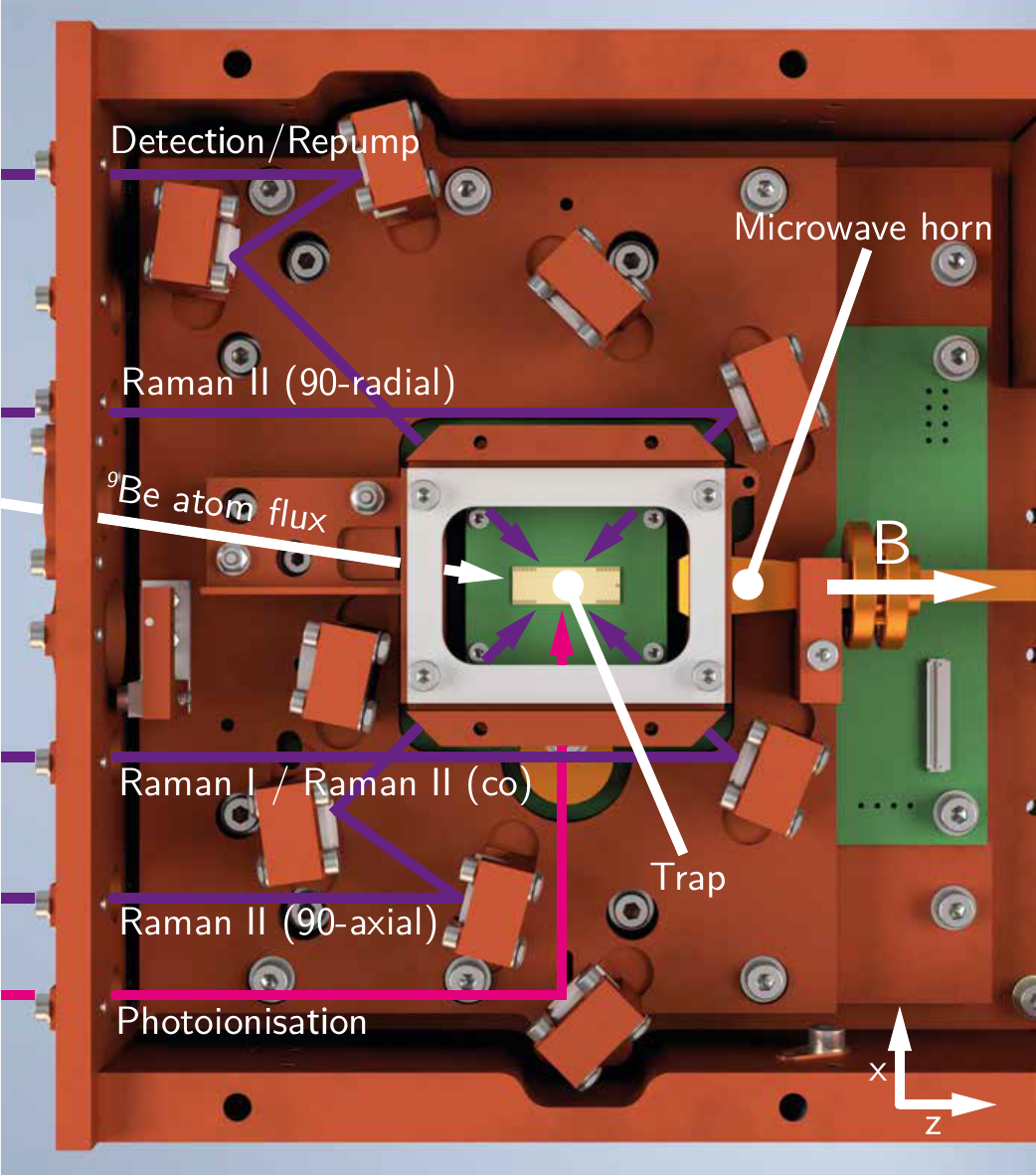}}
\caption{\label{fig:trap_enclosure}\textbf{Trap enclosure.} Top view of the trap enclosure with the Schwarzschild objective and lid removed. The laser beams are directed across the trap chip by an array of mirrors. A horn antenna delivers microwave radiation across the trap.}
\end{figure}

The trap is placed in a cryogenic vacuum apparatus which is inserted into the horizontal magnet bore (Extended Data Figure \ref{fig:apparatus_overview}). The trap enclosure (Extended Data Figure \ref{fig:trap_enclosure}) is surrounded by a vacuum chamber and heat shields, and is held at a temperature of \SI{6.5}{\kelvin}. Laser beams for photoionization, Doppler cooling, repumping, state detection and stimulated Raman transitions are delivered along the magnet bore through a vacuum viewport and are directed across the trap by mirrors mounted within the cryogenic trap enclosure. 

The detection laser beam is delivered at an angle of \SI{45}{\degree} with respect to the magnetic field, such that it has an overlap with all the motional mode eigenvectors. Three configurations of the Raman beams are possible. The Raman I + Raman II (co) beams are co-propagating and produce a negligible wavevector difference, which causes negligible coupling to the motional modes, The other two beam pairs, Raman I + Raman II (90-axial) and Raman I + Raman II (90-radial) have wavevector differences along the axial and the radial motional modes respectively. Fluorescence from the ion is collected by a Schwarzschild objective and directed down the bore of the magnet by a further mirror. Microwave radiation is delivered through a hollow WR-10 waveguide and directed across the trap using a horn antenna. An effusive oven is placed on the room temperature stage of the apparatus and generates a flux of neutral $^9\mathrm{Be}$ atoms, which is directed at the centre of the trap chip.

Ultra-high vacuum at the trap is achieved by cryopumping. Additionally, an activated charcoal getter is added within the trap enclosure to improve the helium partial pressure. Vacuum levels in the room temperature stage of the apparatus are measured to be $<\SI{5e-10}{\milli\bar}$. While we cannot directly measure the vacuum pressure within the trap enclosure, we 
routinely keep ions for several days with primary loss mechanisms related to equipment malfunction or user error, indicating excellent vacuum.

\subsection*{Voltage-Source Detachment}
\begin{figure*}[ht]
\resizebox{\textwidth}{!}{\includegraphics{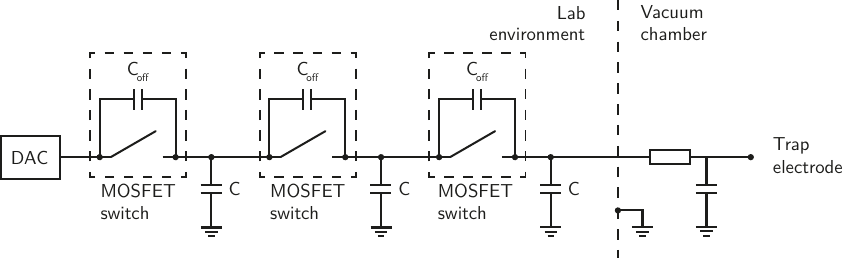}}
\caption{\label{fig:trap_detachment}\textbf{Trap detachment.} Schematic of the switch arrangement used to detach a trap electrode from the external DAC. Three MOSFET-based switches with an on-state resistance of \SI{11}{\ohm} and off-state capacitance $C_\mathrm{off} = \SI{0.45}{\pico\farad}$ are chained and interleaved with capacitors $C = \SI{560}{\pico\farad}$, forming capacitive dividers for improved isolation at radio frequencies. The voltage last applied by the DAC is retained by the combined capacitance of the trap electrode and an RC filter($R = 1~\mathrm{k}\Omega$, $C = 560$~pF) placed within the vacuum apparatus.}
\end{figure*}

As confinement of an ion in a Penning trap requires only static electric and magnetic fields, it is possible to temporarily detach the trap electrodes from the voltage sources external to the Faraday cage formed by the vacuum chamber. In every line connecting a trap electrode to a digital-to-analog converter (DAC) and the lab environment, we place switches which can be actuated using a digital signal (Extended Data Figure \ref{fig:trap_detachment}). In our experimental sequence, we detach the trap after Doppler cooling, which requires the axialization drive to pass to the trap.

The combined capacitance of the trap electrode and the in-vacuum RC filter will retain the voltage that was applied by the DAC before isolating the trap. Although discharging is observed over timescales of minutes, we find no measurable change of ion position or motional mode frequencies during up to \SI{2}{\second} of trap detachment.

To reach maximal isolation from noise present in the lab environment, we concatenate three MOSFET-based switches (G3VM-41QR10, from Omron Electronics). The capacitor $C$ acts as a capacitive divider together with the switch capacitance $C_\mathrm{off}$ to increase isolation performance over the full range of possible motional mode frequencies (dc - \SI{5.118}{\mega\hertz}). 
Simulating this circuit and including parasitic capacitances yields an estimated isolation of 83 dB at dc and 77 dB at \SI{5}{\mega\hertz}.

The option to entirely isolate the trap from the lab environment during parts of the experimental sequence is exclusive to Penning traps. It may permit reduced filtering requirements, in turn enabling faster ion transport, while still allowing low noise levels when the ions are kept at static positions \citeSupp{LincolnLabsTrapDetachment}. It may simplify the task of eliminating external noise sources that are hard to find, or hard to remove.

\bibliographystyleSupp{unsrt}
\bibliographySupp{Supp.bbl}

\section*{Acknowledgments}
This project has received funding from ETH Z\"urich, the ERC under the EU's Horizon 2020 research and innovation programme Grant agreement No. 818195, the EU Quantum Flagship H2020-FETFLAG-2018-03 (Grant Agreement No. 820495 AQTION), and the EU H2020 FET Open project PIEDMONS (Grant No. 801285). S. J. thanks Edgar Brucke for assistance in the cleanroom, and Joseba Alonso Otamendi for his involvements in work building up to the experiment assembly. T. S. thanks Peter Clements for designing the trap detachment PCB. A. B.-S. and C. O. thank the cleanroom staff in particular to T. Weimann, P. Hinze and O. Kerker and acknowledge funding from PTB, QUEST, LUH, and DFG through CRC 1227 DQ-mat, project A01. The authors express their gratitude towards Alfredo Ricci Vasquez and Matteo Simoni for careful reading and critical assessment of the manuscript.
\section*{Author Contributions}
Data taking and analysis were performed by S. J., T. S. and P. H.. The apparatus was primarily built by T. S. and S. J. with contributions from P. H., C. T., C. A., R. O. and M. S.. The trap was fabricated by A. B.-S. with input from C. O..  The manuscript was written by S. J., T. S., P. H and J. H. with input from all authors. The work was supervised by D. K. and J. H. 
\section*{Competing Interests}
A. B.-S. and C. O. are associated with Qudora Technologies, a commercially oriented quantum computing company. The remaining authors declare no other competing interests.
\section*{Data Availability}
The authors declare that the data supporting the findings of this work are available within the paper and the ETH Z\"urich Research Collection repository (open access DOI:10.3929/ethz-b-000627348)
\newpage
\section*{Extended Data}

\end{document}